\documentstyle[11pt]{article}
\newcommand{\blankline}{\vskip .3cm}
\newcommand{\f}{\begin{equation}}
\newcommand{\ff}{\end{equation}}

\begin{document}
\centerline{\LARGE The future of spin networks}
\blankline
\rm
\centerline{Lee Smolin${}^*$}
\blankline
\centerline{\it  Center for Gravitational Physics and Geometry}
\centerline{\it Department of Physics}
 \centerline {\it The Pennsylvania State University}
\centerline{\it University Park, PA, USA 16802}
 \vfill
\centerline{January 26, 1997}
\vfill
\centerline{ABSTRACT}
Since Roger Penrose first introduced the notion of a spin network
as a simple model of discrete quantum geometry, they have 
reappeared in quantum gauge theories, quantum gravity, topological 
quantum field theory and conformal field theory. The  roles that 
spin networks play in these contexts are briefly described, with an 
emphasis on the question of the relationships among them.  It is 
also argued that spin networks and their generalizations provide a 
language which may lead to a unification
of the different approaches to quantum gravity and quantum 
geometry.  This leads to a set of conjectures about the form of
a future theory that may be simultaneously an extension of
the non-perturbative quantization of general relativity and a
non-perturbative formulation of string theory.

\blankline
${}^*$ smolin@phys.psu.edu
\eject

{\it  ``A reformulation is suggested in which quantities normally
requiring continuous coordinates for their description are
eliminated from primary consideration.  In particular, since
space and time have therefore to be eliminated, what might
be called a form of Mach's principle must be invoked: a
relationship of an object to some background space should
not be considered-only the relationships of objects to
each other can have significance." }

	-Roger Penrose,{\it Theory of quantized directions}\cite{1}

\section{Introduction}

Among the many ideas that Roger Penrose has contributed to our 
growing 
understanding of space and time, none is so characteristic as one of
his 
first, which was spin networks\cite{1,sn-roger}.   
Originally invented
as a simple model of a discrete quantum geometry, spin networks
have become
much more than that.  They have been found to provide the
kinematical structure of quantum general 
relativity\cite{volume1,sn1}.
Moreover, they play key roles in 
lattice gauge theory\cite{kogutsusskind,furmanski}
and natural generalizations of them, called 
{\it quantum spin networks}\cite{lou-sn}, 
play an important role in topological field theory and conformal
field theory\cite{witten-cs,tv,louiscft,verlinde,ms,rt,mr}, as well as in 
quantum 
gravity with a 
cosmological constant\cite{sethlee,rsl}.  The best tribute 
I could think of, then, to 
the contribution of Roger Penrose, is to give a quick survey of the
role that spin networks have played in these different
developments in quantum gravity and mathematics. Moreover, this
gives me an opportunity, not only to talk about the role
spin networks have played up till now, but to offer some
conjectures about about how
spin networks might play a key role in bringing together the
most promising developments in quantum gravity of the last
decade, concerned with string theory, non-perturbative quantum
gravity, topological quantum field theory 
and black hole thermodynamics.

I hope the reader will not mind if I tell this story from
a personal point of view.    There seems no better way to
explain the influence that Roger's ideas  have had on those
of us who have been trying to follow his path to an understanding
of the ground on which space, time and the quantum 
are one.

\section{Penrose's spin networks}

The motivation behind spin networks is described well in the
quote above, which comes from an unpublished manuscript
written at the Battelle Institute\cite{1}.  According
to that paper, one wants a description of
quantum geometry that is at the same time:

\begin{itemize}

\item{}Discrete, built purely from combinatorial structures, and

\item{}purely relational, so that it makes reference to no
background notions of space, time or 
geometry\footnote{The roots of this are of course Leibniz's
relational philosophy of space and time.}.

\end{itemize}

The goal of Penrose's construction was to realize a simple model
of such a system.  What he posited is a system consisting of
a number of ``units", each of which has a total angular momentum.
They interact, in ways that conserve total angular momentum.
Without a background geometry, a particle can only have a
total angular momentum, as there is nothing with respect to
which a direction in space may be defined.  

The system is then described by an arbitrary trivalent graph,
$\Gamma$, whose edges are labeled by integers, corresponding
to twice the total angular momentum.  The nodes describe
interactions at which the units meet.  The only condition
imposed is that at the nodes the
conservation of angular momentum must be satisfied.
As there is as yet neither a concept of space or time, it is
left undefined whether the graph is supposed to correspond to
the state or the history of the system\footnote{Of course, this is not
so different from the situation in quantum cosmology.}.

The spin networks that correspond to the states (or histories)
are networks, $\Gamma$, with 
open ends.  The corresponding states may be denoted $|\Gamma >$.  
To take the norm of such a state, one takes it
and its mirror image, and ties up each pair of corresponding
open ends, forming a closed spin network,
called $\Gamma \# \tilde{\Gamma}$.   There is then a
number which may be assigned to any closed spin network,
which is called its value,  $V$.  The norm is then given by
\f
<\Gamma |\Gamma > = V[ \Gamma  \# \tilde{\Gamma} ]
\label{norm}
\ff

This is the entire theory.
The idea
is that everything else about the system must be deducible from
combinatorial principles from the graph that describes the system,
in particular any quantum probabilities that the theory predicts
must be deduced from this norm \ref{norm}.

Penrose  defines the value $V[\Delta ]$
of a closed spin network $\Delta$ as follows.  Each edge
with label $j$ is decomposed into $j$ lines.  At each node
one then has a number of curves that must be joined,
coming from the edges that come into it. By conservation
of angular momentum it is possible to connect up all the
lines coming into the nodes, without joining any two that
come from the same edge.  In general there will be a number
of ways to do this at each node.  Each such choice results in
a number, $N$,  of closed loops. The value $V[\Delta ]$ is then
defined by the sum,
\f
V[\Delta] =\prod_{edges} {1 \over j!} \sum_{routings} \epsilon
(-2)^{N}
\label{value}
\ff
where $\epsilon$ is a sign which is defined such that two routings
that differ by a crossing of lines in any edge have the
opposite sign.  

One consequence of this definition is that the value is invariant
under all the identities of the theory of recoupling of angular
momentum.  Those identities correspond to certain 
graphical relationships among networks, which may be used
to define objects such as $6j$ symbols completely 
combinatorially.  In practice one uses these identities to
reduce a spin network evaluation to combinations of
$6j$ symbols, which is easier and much less prone to error
than trying to keep track of all the signs and factors in
\ref{value}.

The main result of the theory is that the space of directions in
space can be recovered from a notion of probability based on
the value, in the limit of large spin networks\cite{1,sn-roger}.  One
defines a situation in which the angle between two edges
emerging from a complicated network may be measured.  
I will not give the details here, but the point is that the
quantum mechanical probabilities as a function of an
angle, between the units defined by the two edges is
recovered, in the limit that the spins of the edges are
large and the network is also large.

I will not go into more detail about Penrose's formulation
of spin networks, which is explained in
several places\cite{1,sn-roger,sn1}.
Instead, I close this section by mentioning two easy
generalizations of the concept.   

The first generalization is that we can include also graphs
with nodes of any valence.  In this case, there are additional
labels associated with the nodes.  To see why, note that a trivalent
node can be understood as corresponding to a map from the
representations of $SU(2)$ given by the labels on the edges
incident on it to the identity representation.  For the case of
three representations, that map, when it exists, is unique.
But  there may be more than one invariant map from
a product of four or more representations to the identity
representation.
Each such map is called an intertwiner.   For each choice
of representations $j_1,...,j_N$ there is a finite dimensional 
linear space of such intertwiners, called ${\cal V}_{j_1,...,j_N}$.
Each such higher valent node must then be labeled by a state
from the corresponding space\footnote{One way to
give a basis for the space of intertwiners is to give a
decomposition of the higher valent node in terms of a
trivalent network\cite{sn1}.}.

We thus see that the structure of spin networks is based on the
representation theory of $SU(2)$.  This leads to the second
easy generalization, which is to base the formalism on
the representation theory of any Lie Group $G$.   In this
case a spin network is a graph $\Gamma$, whose edges
are labeled by representations, $r$ of $G$, while its nodes
are labeled by the corresponding intertwiners\cite{baez-sn}.  

In fact, the concept of spin networks 
can be defined beyond the representation
theory of Lie groups.  The reason is that the main thing we
are using from representation theory is that there is an
algebraic structure defined on the representations of a
Lie group, such that the product of any two can be decomposed
into a sum of such representations.  In fact, there are algebras
whose representation theory has this property, which are
not Lie groups.  These more general objects are Hopf algebras
and their representation theory may be described in terms
of the theory of monoidal 
categories\cite{joyalstreet,yetter,resh,cp}\footnote{The
precise definitions depend on exactly which properties
are assumed.}.  Thus, there is a still more general class
of spin networks associated with 
these objects\cite{lou-sn,joyalstreet,yetter,resh,cp}.  I 
will describe these a bit later.

\section{Spin networks in lattice gauge theory}

Roger Penrose went on to invent twistor theory, the singularity
theorems and many other things and his spinnets 
remained a kind of talisman for people who believed in the 
possibility that space and time, or at least space, has an underlying
discrete structure.  But after a bit more than a decade they began
reemerge, first in one context, then another.  The first place they
cropped up is lattice gauge theory, which is also the best place
to begin to explain the role they play in quantum gravity.

Here is a too brief survey of lattice gauge 
theory\cite{kogutsusskind}. One has a graph,
$\Lambda$, which is usually taken to be a regular cubic lattice
of dimension $d$.  However, as it costs nothing to describe the
general case, let us allow $\Lambda$ to be completely general.
The graph has nodes $n_i$ and directed edges $e_{ij}$, which connect
node $n_i$ to node $n_j$. (Not every two nodes are connected,
but this notation is still the most convenient.  The only
awkwardness is if two nodes $n_i$ and $n_j$ are joined by
two or more edges, they may then be denoted 
$e_{ij,1},e_{ij,2}, $etc.)   
One picks a compact Lie group, $G$.
A configuration of the theory then consists of a choice of an
element $g_{ij} \in G$ for each directed edge $e_{ij}$ of $\Lambda$.
There is one restriction, which is that $g_{ji}=g_{ij}^{-1}$ which
is where we use the directness of the graph.

There are two cases of interest.  If we are interested in a path
integral formulation in $d$ dimensional spacetime than the
configurations are histories.  If we are interested in a Hamiltonian
formulation then the collection of all configurations is the 
configuration space.  In either case I will denote it 
${\cal C}=\prod_{ij} G$.  That is, the configuration space consists
of one copy of the group $G$ for each edge of the lattice.  I will
denote a particular configuration just as $g_{ij}$.

Actually this is not the physical configuration space.  What makes
the theory interesting is that there is a gauge invariance which
is defined as follows.  A gauge transformation consists of a
choice of an element $h_i$ for each site of the graph.  A gauge
transformation is then the map
\f
g_{ij} \rightarrow g_{ij}^\prime = h^{-1}_i g_{ij} h_j
\ff
The space of all gauge transformations forms a group under
the natural definition in which products of elements are taken
at each site.  This group is called ${\cal G}$.
The basic postulate of the theory is that all physically
meaningful quantities are invariant under this transformation.
This means that the physical configuration space is
${\cal C}_{gauge} = {\cal C} / {\cal G}$.

From now on in this section I will restrict attention to the 
Hamiltonian form of the theory.  In the quantum
theory states will be functions $\Psi $ on ${\cal C}_{gauge}$.
To make the quantum theory we need an inner product
on ${\cal C}_{gauge}$.  There is a natural inner
product on functions of ${\cal C}$ which is given by
\f
<\Psi |\Phi > = \prod_{ij} \int d\mu(g_{ij}) 
\bar{\Psi}(g_{ij}) \Phi (g_{ij})
\label{ip}
\ff
where $d\mu (g_{ij})$ is the Haar measure of the group.
It is interesting to note that this inner product works for
the gauge invariant states as well.  The reason is that
as the integral over the gauge groups is finite for
compact groups, one can
show that on gauge invariant states \ref{ip} 
gives an inner product.

What does all of this have to do with spin networks?  
A great deal, for they provide a very useful orthonormal
basis of physical states.  To see why, we introduce first the
overcomplete set of states based on loops.  A curve
$\gamma$ in the graph $\Lambda$ is a list of edges 
$e_{i_\alpha j_\alpha}=e_\alpha$ such that 
the each begins on the node
the previous one ends on.  A loop is a closed
curve.   Associated with each loop 
$\gamma= e_1 \cdot e_2 \cdot ...e_N$ with $N$ edges we may
define the Wilson loop as
\f
T[\gamma ] = Tr[\prod_\alpha U(g_\alpha )]
\ff
where the $U(g_\alpha )$ are matrices in 
the fundamental representation
of $G$.

The space of all $T[\gamma ]$ forms an overcomplete
basis of ${\cal C}_{gauge}$.  The basis is overcomplete because
of relationships between traces of products of matrices, which
are called the Mandelstam identities
(See for example, \cite{lp1,ls-review,sn1}.  However a 
complete basis of linearly independent states exists and is
given by the spinnets.  A spinnet $\Gamma$ on a graph
$\Lambda$ is given by a set of curves and
intertwiners.  The curves, $\beta_I$,  
each have unique support, which  means that no two curves in 
the set $\beta_I$ share a common edge.  The curves then meet
at a set of nodes $n_X$, which are a subset of the nodes in
$\Lambda$.  The curves are each labeled by a choice of an
irreducible representation $r_I$.  The nodes at which they meet
are labeled by the intertwiners.  If a node $n_X$ has incident
on it a set of curves, $\beta_A$ then the intertwiner 
$s_X$ is a choice
of invariant map from $\otimes_A r_A $ to the trivial representation.

Given a spinnet $\Gamma$ one can construct a gauge invariant
state.  It is easier to say how than to write it.   On each curve,
labeled by $r$ take the products of the matrices of the group
elements in that representation.  Then, at each node, multiply
by the intertwiner that takes those representations incident on it to
the identity.  The result is a gauge invariant functional of the
group elements $g_{ij}$ and hence an element of 
${\cal C}_{gauge}$.  We may denote it
$|\Gamma >$.  To show that these states are complete
and independent is also straightforward, given the inner
product \ref{ip}.  

Note that these spin network states do not satisfy the
set of identities that come from recoupling theory, which is
the case of the original networks of Penrose.  The reason
is that the states $|\Gamma >$ are not purely combinatorial
objects, they are functions of the configuration
$g_{ij}$.

For the simple case of $SU(2)$ it is straightforward to expand
$\Gamma$ as a product of Wilson loops.  One just writes the
representation matrices in the $r$'th representations as
symmetric products of the fundamental representation matrices
$U(g)_A^{\ \  B}$, where $A,B=0,1$.  Thus, for example, the
spin $3/2$ representation matrix may be written
\f
U(g)_{ABC}^{\ \ \ \ \ \  DEF}= U(g)_A^{\ \  (D} U(g)_B^{\ \  E}
U(g)_C^{\ \  F)}
\ff
The intertwiners are then made out of the appropriate combinations
of $\epsilon^{AB} , \epsilon_{AB}$ and $\delta^A_B$.  There is a
beautiful graphical notation for all this, but it is explained well
in a number of places, so I will not describe it here.

The fact that the spin nets provide this basis is not new, it was
mentioned in the first paper on Hamiltonian lattice gauge
theory\cite{kogutsusskind} and exploited in a 
number of later papers in
lattice gauge theory. (See, for example \cite{furmanski}.)
Of course, the lattice gauge theory papers do not use the
word ``spin network" in this context.   In fact, it was not until much
later that the coincidence of the occurrence of the same structure
in gauge theory and speculations about discrete structure of
spacetime was exploited, although it was certainly known to a
number of people.

\section{Spin networks in non-perturbative quantum gravity}

The story of how spin networks entered quantum gravity is a
tale that illustrates the unity and interconnections that have
existed beneath the surface of contemporary theoretical physics,
despite the unfortunate divisions into subfields and camps.  As
Roger has been involved in more than one of the strands of the 
story, this is a story both about his influence and the universality
of some of the central ideas that have formed our understanding
of gauge fields and gravity.  

The first strand follows Roger's work from spin networks to
twistor theory.  There he discovered a very curious 
fact\footnote{also noticed and exploited by Ted Newman and 
Jerzy Plebanski.}, 
which is
the importance of self-duality for an understanding of the dynamics
of the gravitational field.  For in spite of the fact that the Einstein's
equations are duality invariant, the restriction to either the
self-dual or anti-selfdual sector leads to an exactly solvable system,
whose solutions can be described completely in terms of consistency
conditions for the existence of certain complex manifolds.  
The simplification of field equations to the self-dual sector is
equally profound and important for Yang-Mills theory, and indeed
the self-dual or instanton solutions play an essential role now in
both mathematics and physics.  By another basic and
profound fact all this is related to another of the strands of
Roger's work, which is the expression of general relativity in
terms of spinors.  This is because the duality transformation
in four spacetime dimensions induces chiral transformations
that takes left handed spinors into right handed.  

All of this suggested that the dynamics of general relativity itself
might be simplified were it to be expressed in terms of chiral
variables.  The first concrete realization of this 
emerged in two crucial
papers of Amitaba Sen\cite{amitaba}, in which he 
found that the constraints
of the Hamiltonian formulation of general relativity took very simple
polynomial forms when expressed in terms of the self-dual 
(or left-handed) parts of the connection and curvature.  A number
of us puzzled over that paper, but it took Abhay Ashtekar to realize
the full import of what it implied, which was in fact the 
possibility of a hamiltonian formalism for 
general relativity\cite{abhay1}
in which the configuration variable was exactly the self-dual
part of the spacetime connection, $A_{a}^{AB}$, while the
momentum variable is the frame field $\tilde{E}^{a}_{AB}$.
In this formalism the constraints take the polynomial forms
found by Sen; this then became the basis for much of the revival
of work in quantum gravity.  

Not surprisingly, this led directly to a new understanding of the
self-dual sector.  The already simple formulas of the
canonical theory simplified still further when one restricted to
the self-dual sector by setting $F_{ab}^{AB}$, the curvature of
$A_a^{AB}$, to zero, as was discovered by Ted Jacobson and
developed in \cite{atl} and \cite{masonneuman}.   

The Ashtekar formalism
is sometimes  seen to be primarily 
a development of the Hamiltonian
theory, but it led immediately to a reformulation of the lagrangian 
approach as well.  Indeed, it led to more than one, 
as the first action
principle in terms of self-dual variables\cite{sdaction}  led to the
way to the discovery of formulations in which the metric
does not even appear\cite{cdj}.  

Of course, the Ashtekar formalism had profound implications
for quantum gravity, but to trace these we must return to 
lattice gauge theory.    One of the inventors of lattice gauge
theory was Polyakov, who then went on to try to express 
$QCD$ in terms of loop variables in the 
continuum\cite{polyakov-loops}.
One of the strongest memories I have from graduate school
is a seminar given by Sasha 
Polyakov that he began by announcing his
hope to solve $QCD$ exactly by expressing it purely as
a theory of loops.
A different approach to this idea was also developed by
Migdal and Makenko\cite{migdalmakenko}.  
While Polyakov's hope has not so far
been realized, these papers were 
the inspiration for a number of developments,
no less in quantum gravity than in other areas. 

Among these was an attempt to model quantum
general relativity as a lattice gauge theory\cite{melattice}, in 
which the
spacetime connection played the role of the gauge field.  The idea
of this early lattice formulation of quantum gravity 
was to explore non-perturbative approaches to quantum
gravity.  It was particularly motivated
by conjectures of Wilson\cite{wilson}, Parisi\cite{parisi} and
Weinberg\cite{weinberg}  
that perturbatively non-renormalizable theories might in fact
exist were there to be non-trivial fixed points of their
renormalization groups\footnote{For an attempt to
realize this in quantum gravity, see \cite{fixedpoint}.}. 

These were good ideas, but
at the time they led nowhere.  In fact during the early
80's most of us working in quantum gravity were wasting our
time (or at least spending it poorly, considering what we
might have been doing)  
with various perturbative formulations.    My own return to
non-perturbative quantum gravity came
with work with Louis Crane, most of it 
never published, in
which we tried to develop a background independent form of
string theory based on loops variables, which were dual to either
the spacetime metric or connection.  This work was
inspired primarily 
by the papers of Polyakov and Migdal and Makenko.
It was also motivated by work Crane and I had done 
on quantum gravity on fractal spacetimes, in which we
understood that quantum general relativity might exist
if non-perturbative effects lowered the apparent 
dimensionality of spacetime, as seen by the scaling behavior
of propagators above the Planck scale{ll-fractal}.  
As a result, 
we were looking for a way to describe a non-smooth
structure for quantum geometry, in which the 
effective dimension of space at Planck scales would
appear to be less than three,  in terms of loops.

As soon as the Ashtekar formalism appeared, it was clear this
was the way to realize these ideas.   The thing to do was
to construct some kind of discrete geometry from 
Wilson loops made from the Sen-Ashtekar connection.
First, with Paul Renteln, we made a lattice 
formalism\cite{paullee}.  This has
one big disadvantage, which is that it is impossible to realize
diffeomorphism invariant states on it\cite{paul-diffeos}, 
but as a tool for understanding
both the state space and the action of the hamiltonian, it has
been used to good effect since, most 
of all by Ezawa\cite{ezawa}, 
Loll\cite{renata-lattice} and
Gambini and Pullin\cite{GP-lattice}.    
Then with Ted Jacobson we
began investigating a continuum formalism.  There we had a
wonderful surprise, which is the discovery of an infinite class
of physical states-exact solutions to the 
Hamiltonian constraints\cite{tedlee}.  Moreover,
even more than this, we found that the Hamiltonian constraint
acts in a simple way on states made from Wilson loops, with
an action that is concentrated at points of intersections of the loops.

One key question we faced in this work with Jacobson 
was what was the actual space of states of the theory.  
It was clear that the states on which the Hamiltonian constraint
had a simple action were not Fock states.  This was good as
Fock states depend on a background metric, which doesn't
exist in a non-perturbative formalism.  Even if we could
make sense of it, the background metric would interfere
with the action of the diffeomorphisms of space, which are
supposed to be the gauge group, just as much as
a lattice does.  On the other hand, what
was the alternative?  We knew we didn't want to use a lattice
regularization, and we weren't aware of any other choice besides
the lattice regularization or Fock space to define the space of
states of a quantum field theory.  Something new was needed.
To invent it, we were guided by simple physical
ideas.  First, was the physical picture going back
to Penrose and spin networks, that at the Planck scale the structure
of space and time should be discrete.  This picture had been
reinforced by the renormalization group point of view, which
suggested that to realize the conjecture
that the theory is defined through a non-trivial fixed point,
it was necessary that quantum geometry  be based
on  sets of lower dimensionality below the Planck scale.  

Second, we took over the relationship between
quantization of nonabelian electric flux and Wilson
loops coming from $QCD$. This picture comes from ideas of
Holgar Nielson and others, who  had taught us to think of the vacuum 
of
$QCD$ as something like a dual 
superconductor\cite{dualsuper}.  In a superconductor
the magnetic flux is quantized, so the flux through any surface comes
only in integer units of a quantum of flux.   In confinement we
know that the non-abelian electric flux forms tubes whose energy
is proportional to their length, which is also the case for
the quantized magnetic flux lines in a superconductor.  
It is then natural to think that in
the vacuum the non-abelian electric flux is quantized.

Taken together these ideas suggested that a discrete quantum
geometry might be something like an idealized form
of the $QCD$ vacuum, but without any
background structure, so that the flux associated with the
spacetime connection would be quantized.
To realize this picture, we thought about working, not with
a Fock space, but with a space of states spanned by a basis,
each of which was 
made of finite products of Wilson loops.  Thus, we
considered the kinematical state space space 
${\cal H}_{kinematical}$ spanned by the overcomplete basis,
\f
\Psi_{\gamma_i}[A] = \prod_i T[\gamma , A]
\label{discretestates}
\ff
where $\gamma_i$ is any finite set of loops and
$T[\gamma , A] = TrP e^{\int_\gamma A}$ is the Wilson loop,
or traced holonomy.  

States of the form \ref{discretestates} are exactly those 
in which the non-abelian
electric field flux is quantized.  That is, if we identify the 
operator for non-abelian electric flux through a surface $S$
as $E(S)$, then we have
\f
\hat{E}(S) \Psi_\gamma (A) = Int[\gamma , S]^2 \Psi_\gamma (A)
\ff
in the case that there are only simple intersections of the
loop and the surface, i.e. the loop does not intersect itself at the
surface.  (Here $Int[\gamma , S]$ is the intersection number of
the loop and the surface.)

It was thus natural to think that the discrete states
\ref{discretestates} represent
a discrete geometry.  Furthermore, it was obvious immediately that
if the diffeomorphism constraint could be solved on this space of
states, the resulting set of states would be labeled by 
diffeomorphism classes of loops, which is to say knots, links and,
most generally, networks\cite{lsinbook1}.  Thus, knot theory 
immediately emerged
as being important for understanding the state space of quantum
gravity!

These ideas were later formalized by people more mathematical
than ourselves, in the language of rigorous quantum field 
theory\cite{AI,gangof5,baez-sn}\footnote{For a review of the present 
state of the mathematical development of these ideas, see 
\cite{AAinbook}.  For a demonstration of the equivalence of
the formulation of Ashtekar, Lewandowski, Marlof, 
Mour\~{a}u and Thiemann with the earlier formulation
of \cite{lp1,ls-review,sn1,volume1}, see \cite{roberto1}.}.
But I think it is important to emphasize that the roots of these
constructions were in these simple physical ideas, which 
came from $QCD$, renormalization group arguments, and
speculations about the Planck scale.

In fact, it took several years to realize the whole picture.  
For one thing
one had to give a good definition of the operator 
$\hat{E}(S)$.  Formally it looks like,
\f
\hat{E}(S)= \int_S \sqrt{\tilde{E}^a_i \tilde{E}^b_i n_a n_b}
\label{area}
\ff
where $n_a$ is the unit normal of $S$.  The problem is how to define
the operator product and square root.  To do this one needs a
regularization procedure, and all known regularization procedures
depend on a background metric.  The question is then whether one
can define it through a regularization procedure such that
diffeomorphism invariance is not broken, and the operator
takes diffeomorphism invariant states to diffeomorphism
invariant states.  It took some time before a way to do this was
found\cite{ls-review}.  By this time I had realized that the 
non-abelian electric field flux \ref{area} 
was none other than the area
of the surface $S$\cite{ls-review}.  Thus, in quantum gravity, 
discreteness
of area corresponds exactly to confinement in $QCD$\footnote{It
is very interesting to speculate whether the correct way to
formulate $QCD$ rigorously should not be in terms of
the discrete measure \cite{gangof5} which formalizes the
notion of a discrete state space given in \cite{tedlee}. The
problem is that, before the diffeomorphism invariance is
moded out, the space of states is non-seperable. This corresponds
to an unphysical situation in which any displacement at all of
a loop results in an orthogonal state; as a result the theory has
too many degrees of freedom.  On the other hand, $QCD$,
like quantum gravity, clearly cannot be constructed from 
Fock space.  One interesting conjecture to consider is then that
$QCD$ cannot be defined rigorously without coupling it to 
quantum gravity, so that diffeomorphism invariance reduces the
space of spin network states to a countable basis. }.    

Finally,
if one puts in all the constants, the Wilson loop of the 
Sen-Ashtekar connection is actually
\f
T[\gamma , A] = e^{G \int_\gamma A}
\ff
where $G$ is Newton's constant.  Thus, the quanta of area
is proportional to $\hbar G= l_{Planck}^2$.  

The second thing was to construct the diffeomorphism invariant
states which required the loop representation.  The idea to
do this by changing to a representation in which the states
were functions of loops was due to Carlo Rovelli; once we had the 
idea
it did not take long to work it out\cite{lp1}\footnote{The loop
representation had already been invented for $QCD$ by
Gambini and Trias\cite{lp2}, a lot of time could 
have been saved
had we been aware of it.}.  

Another important operator is the volume of a region of space
$V[R]$.    It was immediately clear from formal expressions
that if it could be defined it would be discrete like area and
the discrete eigenvalue would count something happening
at points where three or more loops meet.   This took a long
time to work out\cite{ls-review}, and it 
was this problem that led to the
introduction of spin networks in quantum gravity.  Of course,
the reason spin networks come in is the same as in lattice
gauge theory, because they give an independent basis
for the states of the form \ref{discretestates}.  The 
construction of these
states is the same as in the lattice case, one just takes
any curve in space rather than just the curves on the lattice.
Of course, this was known, and even mentioned at 
times\cite{ls-review},
but it was not exploited.  The main reason was that, unfortunately,
the discovery of physical states associated with non-intersecting
loops \cite{tedlee,lp1}had pushed the question of 
what happened at the 
intersections into the background-even though it was clear-
and emphasized by several 
people\footnote{Especially Berndt Bruegmann and Jorge Pullin.}, 
that the actions of important
operators including volume, the
extrinsic curvature and the hamiltonian constraint
were concentrated at the intersections.  

In fact Roger had a lot to do with the realization that spin networks
are important for quantum gravity.  I was at a workshop in
Cambridge trying to define and diagonalize the volume operator,
and at some point realized that maybe the diagrammatic techniques
Roger had developed to calculate with spin networks could help.
I went to him and he showed me some tricks, which I used to
find that the trivalent spin networks were eigenstates of the
volume operator.  With Carlo Rovelli we then worked out
the action of all the operators we had on spin network 
states and found that these were even simpler than in terms of loops.  
Unfortunately, in the case of volume we got the eigenvalues
wrong-they are zero for all the trivalent networks, as was pointed
out later by Renata 
Loll\cite{renata-volume}\footnote{And also realized independently
by Georgio Imirzi and Michael Reisenberger.}.  In any case,
we had finally realized that the central kinematical concept
in quantum gravity is that the space of diffeomorphism invariant
states is spanned by a basis in one to one correspondence with
embeddings of spin networks\footnote{There have been
many calculations of the spectra of volume, area and length.
See, for example 
\cite{rsl,dpr,RB,tt-length,tt-volume,renata-lattice,AL1,L1,CRR}.}.

The transformation to the loop representation can be
done directly in the spin network basis\cite{sn1}.  When one
mods out by spatial diffeomorphisms one is left with
a state space ${\cal H}_{diffeo}$ which has an independent
basis in one to one correspondence with diffeomorphism
classes of imbeddings of 
spin-networks\footnote{Note that even though we have
gotten rid of the dependence on connections, the states
still are not equivalent under the recoupling identities
of ordinary spin networks.}, which may be
labeled $|\{\Gamma \} >$.  Once we had this space it
was immediate that there is a space of exact solutions,
given by those spin networks without nodes.  
There are other sets of exact solutions, which
include intersections, some of which have been
known for a long time\cite{tedlee,viqar-int}, others of which  were
found recently by Thiemann\cite{qsdi,qsdii}.  Thus,
it seemed that Polyakov's dream that reducing a theory to
loops leads to its exact solutions, is to some extent realized
in quantum gravity.

Is the expression of quantum gravity in terms
of spin networks an important idea, or just a 
technical convenience?  I
believe it is fundamental, probably even more fundamental than
the idea that the states come from applying a quantization procedure
to the infinite dimensional space of connections modulo
gauge transformations.  There are at least four reasons 
to believe this.
First, we have arrived
at a kinematics for quantum gravity that is discrete and 
combinatorial, and it seems likely that such structures are 
fundamental, while continuum concepts such as connections are
artifacts of the myth that space is continuous.    In fact,
at the level of spatially diffeomorphism invariant states
the connections have completely disappeared.  There
is only a space of states spanned by a basis $|\Gamma >$,
where $\Gamma$ now stands for a diffeomorphism class
of spin networks.  The space of states has a natural
inner product
\f
<\Gamma |\Gamma^\prime > = \delta_{\Gamma \Gamma^\prime}
\ff
At this level all operators are combinatorial and topological,
there is no role for a continuum concept such as a connection.

Second, the diffeomorphism classes of spin networks are
somewhat more complicated than the corresponding 
combinatorial and topological classes.   While it might
seem at first that the diffeomorphism equivalence
classes of networks imbedded in a spatial manifold are
labeled only by their topology and connectivity, when
the nodes have sufficiently high valence this is actually not
the case.  In the case of three dimensions,  nodes with five
or more incident edges require continuous parameters to
label their diffeomorphism classes\cite{carlo-grott}.  
If one really believes
that the theory is derived from a classical theory in the
continuum all these should be included.  But if, on the other hand,
one believes that the fundamental structures are discrete, and
the continuum is only an approximation, one might like to
consider as meaningful only those labels of spin networks
which are combinatorial or topological.  Of course, the 
classes labeled by continuous parameters might be needed
if any physically meaningful operator was known that 
measured those parameters, but so far none is known.   
Moreover, even if such an observable existed, it is likely
it could be expressed as well as a slightly less local operator
without the continuous parameters.  For these reasons it
seems best to consider the theory defined by spin networks
defined only up to combinatorics and 
topology\footnote{A related argument has been 
raised\cite{fm-pc}
concerning even some of the topological information, that
concerned with the embedding of the network in the
spatial manifold.  If the discrete structure is really prior to
the manifold, then perhaps imbedding information ought not
to play a role in the fundamental theory.}.

Third, there are difficulties if we take too seriously the
idea that the description of states and operators in terms of
spin networks is in fact the result of a derivation from 
the continuum theory.   Some of these have to do with 
difficulties of the diffeomorphism regularization procedures
that have, so far, been 
developed
\cite{ls-review,lp1,volume1,ham1,RB,roumen-ham,AL1,L1,qsdi}.  
In all of the proposals
so far made, a background metric is introduced which is used
to parameterize a family of point split operators.  The problem
is to define a diffeomorphism invariant operator, which must
have no dependence on the background metric, in the limit
that the regulator is removed, bringing the operators together.

There is a very nice thing about these constructions, which is
that when they succeed in constructing a diffeomorphism
invariant operator, that operator is necessarily 
finite\cite{ls-review}.  The
reason is that any divergence, if present, is measured in units
of the background metric.  If the result of taking the limit in
which the regulator is removed is an operator that does not
depend on the background metric it cannot be proportional to
any divergent quantity; it must then be finite.  In fact, all
cases that have been worked out go exactly like this, and this
may be counted as one of the successes of non-perturbative
quantum gravity: diffeomorphism invariance is sufficient
to guarantee finiteness of operator products, defined through
such a regularization procedures.

While this works simply for the case of the area operator, two
kinds of problems appear when it is applied to more complicated
cases including the volume, hamiltonian constraint and
hamiltonians\footnote{In the canonical formalism a hamiltonian
is obtained whenever the time part of the gauge 
invariance is fixed.}.  
The first is ambiguity; different regularization procedures result
in different diffeomorphism invariant operators.  This is of
course nothing new, it afflicts all quantum field theories.
The second problem is more serious, it is that in these cases
one must use highly non-trivial operator orderings in order
to achieve a diffeomorphism invariant operator.  For instance,
in the loop or spin network representation the limit is taken
along families of operators that measure various features of
the loops at intersections, beyond that information which is
gathered by those operators that appear in the naive
transcription of the corresponding classical quantity.  There
can be no objection, at least in the loop representation to
the insertion of such operators, but it makes the constructions
highly non-trivial\footnote{In the connection representation
\cite{} the situation is not as good because the additional
operator dependence needed cannot be expressed in terms
of the basic operators involving the connection (and not 
the loops directly) without additional operator products which
themselves need regularization.  So it is not clear that an
honest point split regularization can be achieved in the
connection representation;
one may then have no resort but to invent a category
of ``state dependent" regularization procedures.  This problem
does not occur in the loop representation because
there one can construct completely well defined local operators
that measure the support of the loop directly, such as
$\hat{\dot{\gamma}}^a (x) |\gamma > \equiv 
\int ds \dot{\gamma}^a (s) \delta^3(\gamma (s),x) |\gamma >$.}.
As a result
it is far from clear what real advantage comes from
taking seriously the program of deriving the quantum theory
from the classical theory, especially as the quantum theory is 
believed in reality to be the exact description, while the
classical description should be only an approximation.  It is
as if one tried to derive Newtonian mechanics by a systematic
procedure from Ptolemy's astronomy.

Yet another problem is that all forms of the Hamiltonian
constraint so far developed have a problem with the
continuum limit, in that the physical degrees of freedom
are too localized in finite regions of networks, and do not
propagate in a way that can lead to long range correlations
in a continuum limit\cite{instability}.

The last reason to take the spin network description as
more fundamental than the classical connections 
is that $SU(2)$ spin 
networks immediately generalize to a large class of cases which
furthermore, makes contact with conformal field theory,
topological field theory and, through them, to string theory.
Furthermore, the cases that make contact with conformal
field theory are necessarily related to quantum groups which
do not correspond in any exact sense to classical connections.
I now turn to this story, which is another tale of how spin networks
came into mathematics and physics.

\section{Spin networks in topological quantum field theory and
conformal field theory}

Topological quantum field theory can be approached from three
different directions, the purely combinatorial\cite{lou-sn}, 
the category
theoretical\cite{tv,rt,louiscft} and through a path integral 
formulation of a
quantum field theory\cite{witten-cs}.  Spin 
networks enter into all three of
these.  I will sketch briefly how each approach works with
respect to the best studied example of a quantum field theory,
which is Chern-Simons theory.

We begin with the path integral definition of quantum Chern-Simons
theory\cite{witten-cs}, which is given by,
\f
Z= \int d\mu (A) e^{{\imath k \over 4\pi} S^{CS}(A)}  ,
\label{chernsimon}
\ff
where $S= \int_\Sigma Tr(A\wedge dA + {2\over 3} A \wedge A 
\wedge A)$ 
is the Chern-Simons action on a compact three
manifold $\Sigma$
and $A$ is a connection one form for a gauge group $G$.
We note that the action $S$ is invariant under small
gauge transformations but transforms under large gauge 
transformations
as $S \rightarrow S^\prime = S + 8 \pi^2 n$, where $n$ is an integral
winding number.  As a result, the coupling constant 
$k$ must be an integer so that the theory is invariant
under large gauge transformations. 

At the formal level, the quantum field theory defined by 
\ref{chernsimon} is
diffeomorphism invariant.  In fact, the theory can be defined
so that this is the case,  although this has not been done, at
least so far, by defining an honest diffeomorphism invariant
measure $d\mu (A) $ in \ref{chernsimon} with 
respect to which $expiS_{CS}$
is measurable\footnote{Note that $expiS_{CS}$ is
{\it not} measurable by the kinds of measures described
in \cite{gangof5,AAinbook}.}.  Let me put this 
very interesting question to one side
and simply describe the result.  

As the theory is diffeomorphism invariant, the expectation values
of products of local operators do not contain very much information.
Instead, the theory becomes interesting if one studies the
expectation values of non-local operators.  For example, if
one has a loop, $\gamma$, then one can compute
\f
< T[\gamma ] > \equiv  K^k[\gamma ] ={1 \over Z} 
\int d\mu (A) e^{{k \over 4\pi} S^{CS}(A)} T[\gamma ,A]
\label{csloop}
\ff
By inspection $K^k[\gamma ]$ must be a knot invariant, in fact, as 
Witten
discovered in a justly celebrated work, it is equal to a very
important invariant, which is the Kauffman bracket. This invariant
associates to every knot a function of $k$.   However, there
is an important subtlety.  The integral  
\ref{csloop} has divergences, which
require that the operator products in the definition of the Wilson
loop be regularized.  This is done by smearing the loop into a 
strip and then taking the limit in which the width of the strip
is taken to zero.  This introduces additional degrees of freedom
associated with the winding of the strip.  We describe this as
saying that the Kauffman bracket is really an invariant of
a ``framed" loop, or strip.

In fact, one can use Chern-Simons theory to give an expectation
value to any spin-network $\Gamma$.  Given a spin network 
$\Gamma$,
there is a gauge invariant 
functional of $A_a$ associated to it, given by the
continuum version of the procedure I
described in section 3, called $T[\Gamma , A]$.  One way
to express this is just to write out the spin network as a sum of
products of loops, as I described in section 2, 
and then $T[\Gamma , A]$ is the corresponding
sum of products of Wilson loops.

One then computes
\f
< T[\Gamma ] > = K^k [\Gamma ] ={1 \over Z} 
\int d\mu (A) e^{{k \over 4\pi} S^{CS}(A)} T[\Gamma ,A]
\label{csnet}
\ff

This turns out to be almost, but not quite, an 
invariant of spin networks.  To define it we must deal with 
the problem of framing just discussed.
The integral \ref{csnet}, when regularized, as it must be, will
give an invariant of framed spin networks.
The problem is to do the regularization such that a
consistent definition of a framed spin network results.  It turns out
that there is a very beautiful way to do this, which is called
a quantum spin network\cite{lou-sn}.  
These are a family of deformations of
Penrose's spin networks, which are labeled by a deformation 
parameter
$q$.  In the case of Chern-Simons theory $q= e^{4\pi/(k+2)}$.   

Quantum spin networks may be understood as
built up from the representation theory of
quantum groups, which are deformations of Lie
algebras.  Quantum groups are Hopf algebras, but they do not
correspond to groups, thus the quantum spin networks
are in the category of extensions of spin networks that
do not correspond to gauge invariant states of classical
connections.  They 
differ from ordinary spin networks in 
several ways.  First, they satisfy a modified 
set of recoupling identities, which
define a set of {\it quantum} $6j$ symbols that depend on $q$.
Second, the possible spins on the edges cannot be greater than
$k+1$.  Third,  unlike Penrose's formula for the value of
a spin network, their
invariants can detect information about the imbedding of the 
network
in the spatial manifold.  Because they can detect twisting they are
even chiral, invariants such as the Kauffman invariant
can tell chiral knots from their mirror images.  

In the limit $q \rightarrow 1$, which corresponds to the
classical limit $k \rightarrow \infty$ of Chern-Simons theory, the 
Kauffman bracket
$K^k[\Gamma ]$ becomes proportional to the 
Penrose evaluation $P[\Gamma ]$ \cite{lou-sn}.

The Kauffman bracket can in fact be defined 
combinatorially\cite{lou-sn}, 
independently of the quantum field theory defined by 
\ref{csnet}.  
Thus, while there is no measure by which 
\ref{csnet} is known to be
defined, all expectation values of the form \ref{csnet} 
can be computed
in closed form.

Besides the path integral and the combinatorial approaches, there
is a third framework within which Chern-Simons theory may be 
understood.
This is the axiomatic, or category theoretic approach initiated by
Segal\cite{segal} and Atiyah\cite{atiyah}.  
This approach begins by choosing an arbitrary
closed two surface $S$ in the spatial manifold 
$\Sigma$ that splits it into two 
halves, each
with boundary $S$.  For a given evaluation 
\ref{csnet}, $S$ has on it a number
of ``punctures", which are the points where the edges of $\Gamma$
intersect $S$.  The punctures are labeled by spins, $j$, which are
the labels of the corresponding edges.  

To each closed surface $S$, with labeled punctures $j_\alpha$, one
then associates a finite dimensional Hilbert space,
${\cal H}_{S,j_\alpha}$.  This may be constructed through the
canonical quantization of the Chern-Simons theory, but it may also
be constructed combinatorially. One then
constructs a topological quantum field theory for the manifold
$\Sigma$, not by constructing one Hilbert space but by constructing
a whole family of related Hilbert spaces, one for each punctured,
labeled surface $S$ that splits $\Sigma$ into two parts.

The content of a topological field theory is in relationships
that are defined between these Hilbert space ${\cal H}_{S,j_\alpha}$.
The theory has to do with topology because these relationships
correspond to basic topological operations.  For example, 
the operation of reversal of orientation of $S$ is associated to the
hermitian conjugate in ${\cal H}_{S,j_\alpha}$.  Then suppose
one has a cobordism ${\cal C}=\{ \rho, \Gamma \}$ between 
two punctured, labeled surfaces, $S,j_\alpha$ and $S^\prime , 
j_\alpha^\prime$.
This consists of a three manifold $\rho$ with boundary 
$\partial \rho = S \cup S^\prime$ together with an embedded spin 
network
$\Gamma$ that meets the boundary at the punctures, such that the 
labels
agree.  Then to $\cal C$ there corresponds a linear map
\f
{\cal M}_{\cal C} : {\cal H}_{S,j_\alpha } \rightarrow 
{\cal H}_{S^\prime ,j_\alpha^\prime }
\ff
In particular, suppose one boundary is trivial, so that 
$\partial \rho = S $.  Then the pair $\rho , \Gamma$ must induce
a state
\f
|\rho , \Gamma > \in {\cal H}_{S,j_\alpha } 
\ff

This is actually very much like Penrose's original notion, in which
states are associated with open spin networks.  However, these are
more subtle, as the phase of the state depends on the 
fact that the open edges end on a compact two dimensional
surface.  In fact, the framing dependence of
the quantum spin networks implies that  large diffeomorphisms of 
the two surface
can change the phase of the state.   

One then constructs invariants of imbedings of (quantum)
spin networks in compact three manifolds from the inner product
of the topological quantum field theory.  Given $\Gamma$ imbedded
in $\Sigma$, one splits them along any surface $S$, giving rise to
two manifolds $\Sigma_{1,2}$ each with boundary $S$. In each half
there is an embedded spin network $\Gamma_{1,2}$, each of which 
meets
$S$ at the same set of punctures $j_\alpha$.  One then has two
states, which we may call $|1>$ and $|2>$ in ${\cal H}_{S,j_\alpha}$.
The Kauffman bracket is then given by
\f
K^k[\Sigma, \Gamma ] = N <1|2>.
\ff
where $N$ is a normalization factor, and I have included explicitly
the dependence on the manifold topology.
The identities of the topological quantum field theory then guarantee
that this is independent of the way the surface $S$ that splits 
$\Sigma$ into two halves is chosen.

This is again a generalization of the notion of Penrose, as
$K^k[\Sigma, \Gamma ] $ is a deformation of Penrose's
value $V[\Gamma ]$.  It is just that the
newer notion is more powerful, as it can measure
features of the topology of $\Sigma$ and the 
imbedding.  But, in both cases, it is the recoupling
identities from the representation theory that guarantee
that the inner product is independent of how the surface
$S$ is drawn that splits a closed network and manifold into
two open halves.  It is this relationship between topology and
representation theory that underlies the categorical
approach to topological quantum field theory.  

Another aspect of this construction is that it is related to
conformal field theories.
The Hilbert spaces ${\cal H}_{S,j_\alpha}$ are exactly the spaces
of conformal blocks of conformal field theories 
defined on $S$ \cite{witten-cs,louiscft,mr}.
This circumstance reflects a deep mathematical relationship
between the representation theory of a quantum groups $G_q$ at 
roots of
unity and the representation theory of the corresponding loop
group $\hat{G}$ at level $k$\cite{ms,mr,kashdan}.  

\section{Spin networks as a bridge between quantum gravity
and conformal field theory}

We have mentioned four different contexts in which spin networks 
or
their more elegant cousins, quantum spin networks, appear: 
Penrose's
original formulation of a discrete angular geometry, lattice gauge
theory, non-perturbative quantum gravity and topological quantum
field theory and conformal field theory.  It may of course be that
this is just a kind of coincidence with nothing deep attached to it;
the fact that the group $SU(2)$ describes both spin and isospin is
usually thought to be genuine coincidence.  On the other hand, 
it is worth contemplating the possibility that rather than being
a coincidence, this is a clue that spin networks, or something like
them, are part of the proper language for describing the geometry
of space and time at the Planck scale.  

One reason for thinking this is that the Chern-Simons invariant
plays a big role in quantum general relativity itself.  The reason
is that when the cosmological constant, $\Lambda$, is non-zero,
there is an exact physical state of quantum general relativity
given by\cite{kodama}
\f
\Psi_{CS}(A) = e^{{k\over 4\pi }S^{CS}(A)}
\label{csstate}
\ff
where $k$ is related to the Newton's and cosmological
constants by, 
\f
G^2 \Lambda = {6\pi \over k}
\label{magic}
\ff
By \ref{csnet} in the loop representation, this state is
exactly given by the Kauffman 
bracket $K^k[\Gamma ]$\cite{BGP,GP}.  
Furthermore, this state has a good classical 
limit\cite{chopinlee},
in the limit of large $k$ or small $\Lambda$, which is
De Sitter spacetime.  In fact, this is the only of the
exact states of quantum gravity that is known to have
both an exact
description in terms of spin networks and a good
classical limit.  The fact that it is related closely to
topological quantum field theory and conformal field
theory is unlikely to be coincidence.

But perhaps the best reason for thinking that
spin networks might be fundamental for quantum geometry
are that they are 
part of a cluster of mathematical structures that connect algebra,
representation theory and topology.  We have
already seen the beginning of this at the end of the
last section, but the relationship goes still 
deeper.  The core of it is
a set of deep relationships between the representation theories
of various algebras and topological problems in various dimensions,
which are beautifully expressed in the language of the theory of
tensor categories\cite{joyalstreet,yetter,resh,cp}.  I believe it is very 
possible that it is there, in
that locus of representation theory and topology, that the
fundamental quantum structures that underlie spacetime 
are to be found.

If this is so then there should be contexts in which physical
relationships can be found between structures found in quantum
gravity and those in topological field theory and conformal
field theory.  To look for them,  we may try to find
contexts in which we can use structures from conformal
field theory to solve some problem in 
quantum gravity.  There are in fact two contexts in which exactly
this can be done.  In both of them we consider quantum gravity in a
region of spacetime, $M$, surrounded by a boundary, $\partial M= 
S\times R$,
on which a particular boundary condition has been imposed, which is
\f
e^i={k G^2  \over 2 \pi} f^i
\label{sdbc}
\ff
where $e^i$ is the pull back of the self-dual two form of the metric
to the boundary and $f^i$ is the pullback of the self-dual part of
the curvature.
These are called self-dual boundary conditions.

There are two interesting cases in which these boundary conditions 
may
be realized.  The first is in the case of Euclidean quantum gravity
with a cosmological constant $\Lambda$,  which
are related by \ref{magic} \cite{boundary}

The second, and perhaps more exciting case, is Minkowskian 
quantum
gravity, where we require that the geometry of the boundary may
be matched to a static spherical surface in the
classical Schwarzchild solution of radius $R$, in 
which case\cite{kirill1} 
\f
k={A[S] \over 2 }{R \over 2M} = { \pi R^3 \over M}
\ff
where $A[S] $ is the area of the surface.  

In each of these two cases there is an algebra of observables defined
on the surface $S$, which we may call ${\cal A}_S$.  What is very
interesting is that as long as the boundary conditions are
chosen so that $k$ is an integer, a representation can be constructed 
for
this algebra from direct sums of the 
finite dimensional Hilbert spaces ${\cal H}_{S, j_\alpha}$,
which are the spaces of conformal blocks associated with Chern-
Simons
theory at level $k$.  I will describe how this goes in the case of 
Euclidean
quantum gravity\cite{boundary}, the story is similar for the 
Minkowskian 
black hole case\cite{kirill1}.

Among ${\cal A}_S$ are observables $A[R]$ which measure the areas 
of
every region $R$ of $S$.  The eigenspaces of these are associated 
with
punctures on the surface, labeled by spins $j_\alpha$.  Each
set of punctures and labels corresponds to  areas
\f
A[R]= {l_{Planck}^2 \over 2} \sum_{\alpha \in R} \sqrt{j_\alpha 
(j_\alpha +1)}
\ff
Associated to each set of punctures and representations are two state
spaces ${\cal H}_{S, j_\alpha}$ from conformal field theory and
${\cal H}_{S, j_\alpha}^{QG}$, which is a subspace of 
the space of states of quantum gravity, ${\cal H}_{S, j_\alpha}^{QG}$
 with cosmological constant \ref{magic}
consisting of all (quantum) spin networks that enter the boundary
at the punctures with edges given by the spins of the punctures.
This is the diffeomorphism invariant subspace, in which we have
applied the diffeomorphisms in the interior of the surface, but not
yet the Hamiltonian constraint.  What is very interesting is that there
is a space of physical states of quantum gravity that lives
inside each of these which is isomorphic
to the space ${\cal H}_{S, j_\alpha}$.  These states are related to
the Chern-Simons state \ref{csstate}.  

There are two remarkable things about these physical states of 
quantum
gravity. The first is that all the operators in the surface algebra
${\cal A}_S$ of quantum gravity 
can be represented directly in terms of operators that
either act within each space of conformal blocks or map between 
them.
The second thing is that in the limit
$k \rightarrow \infty$ 
the dimensionality of these state spaces saturate
the Bekenstein bound\cite{boundary},
\f
Dim {\cal H}_{S, j_\alpha} = e^{c A_{S, j_\alpha} / l_{Planck}^2}
\ff
where $A_{S, j_\alpha}$ is the eigenvalue for the area of the surface
coming from quantum gravity, and $c$ is a dimensionless constant
of order unity (which is not 
equal to $1/4$\footnote{We do not expect to get the $1/4$ right.
Among other reasons, we expect there must be a finite 
renormalization
of Newton's constant between its bare, Planck scale value and the
measured macroscopic value.}.)

This formula is remarkable as it involves input from both conformal
field theory and quantum gravity, and it supports a conjecture
coming from black hole thermodynamics.  In my opinion it is 
evidence
for the existence of a non-trivial connection between these things.

Given that there are independent arguments for the Bekenstein
bound, we may conjecture that these states are all the exact 
physical states in the theory with the boundary 
conditions \ref{sdbc}.
In this case we may write a formula for the physical state space
of quantum gravity with the self-dual boundary 
conditions \ref{sdbc}
\f
{\cal H}_S^{physical} = \sum_{j_\alpha}{\cal H}_{S, j_\alpha}
\ff
expressing the physical state space of quantum gravity in 
$3+1$ dimensions as a direct sum of spaces of conformal blocks
associated with Chern-Simons theory.

\section{The dynamics of spin networks}

The main result of non-perturbative quantum gravity so far is that
the kinematics of the spacetime are described in terms of a basis
of states which are in one to one correspondence with embeddings
of spin networks.  The next step is to study dynamics.  Three
approaches to the dynamics of quantum gravity are being studied 
which
employ spin networks.  The first is to express the Hamiltonian
constraint as an operator on spin 
networks\cite{lp1,ham1,RB,roumen-ham}.   
The main technical problem
here is to find a regularization and renormalization of the 
Hamiltonian constraint so that a space of solutions, which are
exact physical states, may be constructed.  There are several
different approaches that have been pursued, the
state of the art is presently represented by the recent papers
of Thiemann\cite{qsdi,qsdii} and 
Borissov\cite{roumen-ham}.    As I have
already mentioned, there are several kinds of exact solutions
known, which follow from different approaches to defining
the quantum Hamiltonian constraint.

The second approach is to fix the time gauge, so that the dynamics
is generated by a Hamiltonian rather than a Hamiltonian constraint.
Several cases are under study, in which time is taken to be
given by the value of a matter degree of
freedom, such as a scalar field\cite{meindieter,ham1,roumen-ham} 
or 
dust\cite{brownkuchar}.  One may also
try to employ an intrinsic notion of 
time such as that based on
the Chern-Simons invariant of the Ashtekar-Sen 
connection\cite{chopinlee} or one derived from
an abeleanization of the constraints\cite{fm1}.

The third approach is to write the time evolution operator directly
by means of a path integral.  In such a formulation, the continuous
path integrals of the formal theory are replaced by sums over four
dimensional spin networks, each of which represents a discrete 
spacetime geometry.  Although this is the newest approach to 
dynamics,
there are three approaches being pursued at the present time.
The first  is
the effort to extend topological field theory from three to
four dimensions\cite{4dtqft}.  The second is recent work of
Reisenberger and Rovelli in which a time evolution operator
for Euclidean quantum gravity is found\cite{RR}.  
There are very interesting
similarities between the results of these two programs, which 
suggest
either that quantum gravity will be a four dimensional topological
quantum field theory, or will be closely related to 
it\footnote{How could quantum gravity be a topological quantum
field theory, if those theories have only finite numbers of degrees
of freedom, associated with surfaces?  The answer is the 
Bekenstein bound\cite{bekensteinbound}, which 
tells us that any subspace of
the state space of quantum gravity associated with measurements 
made
in a bounded region with finite surface area must be finite.  The
holographic hypothesis of 
'tHooft\cite{thooft} and Susskind\cite{lenny-holo} 
then conjectures that
such a theory is defined in terms of state spaces and observables
on surfaces.  As far as I know, the only consistent non-perturbative
quantum field theories that realize the holographic hypothesis are
topological quantum field theories. }

However, both of these formulations are Euclidean. One might prefer 
to
construct a path integral for the Minkowskian signature theory
directly.  If this were possible one could 
implement the causal structure
directly at the level of the four dimensional networks that provide
the histories for spin networks.  It turns out that exactly this can
be accomplished, leading to a class of path integrals for the 
evolution of spin networks in Minkowskian time, in which amplitude
is expressed in terms of a discrete structure that is both a
four dimensional spin network and a causal set 
\cite{fotinilee}\footnote{That is a
graph on which there are a set of causal relations such as one
finds amongst points in Minkowskian spacetime as in
\cite{causalset}.}.

\section{The future}

We have seen that Penrose's original intuition has been confirmed:
spin networks do provide a kinematical framework for 
understanding
quantum geometry.  At least they do if by quantum geometry we
mean the quantization of Einstein's classical theory.  However, there
are many reasons to believe that in reality quantum geometry is a
good deal subtler than that.  Among these are,

\begin{itemize}

\item{}There are other degrees of freedom besides the spacetime
metric, and we wonder if it might be possible 
to understand them in some framework
that unifies them with geometry.  Certainly string theory provides
evidence from the perturbative level that this should be possible.

\item{}In any quantum theory of gravity based on some
discrete framework that has critical behavior, leading to a
classical limit in which the  universe grows very large
in Planck units, the effective action that governs the large
scale dynamics of spacetime should be, to good approximation,
described by general relativity.  (This is an old argument,
based on the usual renormalization group considerations.)
Thus, there is no reason to suppose that the microscopic 
dynamics has anything to do with general relativity.  Instead,
the problem is to show that the theory does have
critical behavior necessary for the discrete universes it 
describes to grow big and classical.  

\item{}There are a number of problems that we have no idea how
to solve given the ideas and structures we now have, which include
the problem of the smallness of the cosmological constant, the other
problems of the specialness of the parameters of low energy
physics and cosmology, the black hole information
puzzle, as well as the interpretational problems
of quantum cosmology.  The solutions to these problems might
very well require new mathematical structures, we should thus
be looking for them.

\end{itemize}

I would then like to close with a list of conjectures 
about what spin networks might have to do with a form
of quantum gravity that would address these issues.

\begin{itemize}

\item There is a quantum theory X, defined in a purely
algebraic fashion, without respect to any background manifold or
geometry, which has as a classical limit $3+1$
general relativity coupled to certain matter fields.

\item The perturbative theory around the classical limit of
theory X is described by some perturbative string theory.

\item  The kinematical structure of X is defined from the
representation theory of some Hopf algebra, associated with
the groups which play special roles in perturbative string theory,
such as $SO(8)$ and the exceptional groups.  The natural language
for theory X will then be that of tensor categories.  As such,
a generalization of spin networks, such as one finds in 
category theory\cite{joyalstreet,yetter,cp} will play a 
role in the theory.

\item  Theory X will realize directly the holographic hypothesis
and the Bekenstein bound, because it will be interpretable in terms
of state spaces and observable algebras associated with boundaries
that divide the universe into parts, following the categorical
framework for topological quantum field theory.  Thus, it will
resolve the interpretational problems of quantum cosmology
along the lines of the proposals of Crane\cite{louis-qc}, 
Rovelli\cite{carlo-relational} and the 
author\cite{pluralistic,book}.

\item Geometry will arise from theory X when a system has a critical
behavior, as in the case of random surface theory and models of
quantum gravity based on dynamical triangulation and Regge 
calculus.
When they are so defined, operators that measure geometrical
quantities will have discrete spectra, as in the case of areas and
volumes in quantum general relativity.  Further, there will be
algebraic conditions required for the theory to be critical at the
non-perturbative level, which  will be related to the conditions that 
are
required so that a conformal field theory may describe a 
perturbative
string theory.

\item As there is no agent external to the universe to tune
some relevant coupling to make the system critical, the critical
behavior of cosmologies defined by Theory X, necessary for the
universe to get big and classical, must be the result of some
mechanism of self-organized criticality.  We may conjecture
that self-organized critical behavior corresponds to Minkowskian
signature quantum gravity in the same sense that equilibrium
critical behavior corresponds to Euclidean quantum field theory.

\end{itemize}

This last point takes us beyond what has been mentioned here, the
rational behind it is described in a companion paper\cite{origins}.  
The other points come from taking an optimistic stance,
in which one assumes that the main robust results of the
different approaches to quantum gravity are all true.  Thus, rather
than seeing string theory and non-perturbative quantum gravity as
somehow opposing each other, I think it is more fruitful to believe
that they represent different regimes of the same fundamental 
theory.
If this is true, we may be closer to that theory than we think.

\section*{ACKNOWLEDGEMENTS}

It is a pleasure to thank  John Baez, Louis Crane, 
Ted Jacobson, Louis Kauffman, Carlo Rovelli, Roumen
Borissov, Seth Major and, or course, Roger Penrose, for teaching
me most of what I know about spin networks (as well as a great deal
more that I forgot.)  I am indebted also to Roumen Borissov,
Kirill Krasnov and Fotini Markopoulou for many conversations
about the future direction of this subject as well as for
comments on a draft of this paper.  I would also
like to thank Nick Khuri and Daniel Amati for hospitality at
The Rockefeller University and SISSA during the course
of this work.  This work
was supported by the NSF grant  PHY-93-96246
to The Pennsylvania State University.


\begin{thebibliography}{99}

\bibitem{1}R. Penrose, {\it Theory of quantized directions}
unpublished manuscript.

\bibitem{sn-roger}R Penrose: in {\it Quantum theory and 
beyond}  ed T Bastin, Cambridge U Press 1971;
in {\it Advances in Twistor Theory}, ed. L. P. Hughston and R. S. 
Ward,
(Pitman,1979) p. 301; in {\it Combinatorial Mathematics and
its Application} (ed. D. J. A. Welsh) (Academic Press,1971).


\bibitem{volume1}C. Rovelli and L. Smolin
{\it Discreteness of area and volume in quantum gravity}
 Nuclear Physics B 442 (1995) 593.  Erratum: Nucl. Phys.
B 456 (1995) 734.

\bibitem{sn1}C. Rovelli and L. Smolin,  
``Spin networks and quantum gravity"  
gr-qc/9505006, Physical Review  

\bibitem{kogutsusskind}J. Kogut and L. Susskind,  Phys. Rev.
D 11 (1975) 395.

\bibitem{furmanski} W. Furmanski and A. Kowala, Nucl. Phys.  B
291 (1987) 594.


\bibitem{lou-sn}L. Kauffman and S. Lins {\it Tempereley-Lieb
Recoupling Theory and Invariants of 3-Manifolds}
Princeton University Press, 1994, and references therein.

\bibitem{witten-cs}E. Witten, {\it Quantum field theory
and the Jones Polynomial} Commun. Math. Phys.
121 (1989) 351.

\bibitem{tv} V Turaev O Viro: Topology 31 (1992) 865

\bibitem{louiscft}L. Crane, Commun. Math. Phys. 135 (1991) 615;
Phys. Lett. B 259 (1991) 243.

\bibitem{verlinde}E. Verlinde, {\it Fusion rules and modular
transformations in 2D conformal field theory}
Nucl. Phys. B 300 (1988) 360.

\bibitem{ms}G. Moore and N. Seiberg, {\it Classical and
quantum conformal field theories} Comun. Math. Phys.
123 (1988) 177.

\bibitem{mr}G. Moore and N. Yu. Reshetikhin, {\it A comment
on quantum group symmetry in conformal field theory},
Nucl. Phys. B 328 (1989) 557.

\bibitem{rt}N. Yu. Reshetikhin and V. G. Turaev, 
{\it Ribbon graphs and their invariants derived from quantum
groups}, Commun. Math. Phys. 127 (1990) 1;
{\it Invariants of 3-manifolds via link polynomials and
quantum groups}, Invent. Math. 103 (1991) 547.

 \bibitem{sethlee}S. Major and L. Smolin, {\it Quantum deformation
of quantum gravity} gr-qc/9512020, Nucl. Phys. B 473
(1996) 267.

\bibitem{rsl}R. Borissov, S. Major and L. Smolin, {\it The
geometry of quantum spin networks} gr-qc/9512043,
Class. and Quant. Grav. 12 (1996) 3183.

\bibitem{baez-sn}J. Baez, {\it Spin network states
in gauge theory}, gr-qc/941107, Adv. Math.
(1995), to appear; {\it Spin networks in non-perturbative
quantum gravity}, gr-qc/9504036.

\bibitem{joyalstreet}A. Joyal and R. Street, 
{\it Braided monoidal categories}, Macqurie Mathematics
Reports, no. 860081 (1986); {\it The geometry of
tensor calculus I}Adv. Math. 88 (1991) 55.

\bibitem{yetter}D. N. Yetter, {\it Quantum groups and
representations of monoidal categories}, Math.
Proc. Cam. Phil. Soc.  108 (1990) 261.

\bibitem{resh}N. Yu. Reshetikhin, {\it Quasitriangular
Hopf algebras and invariants of tangles}, Lenningrad
Math. J. 1 (1990) 491.

\bibitem{cp}V. Chari and A. Pressley, {\it A
guide to Quantum Groups}, Cambridge University Press, 1994.
For the generalization of spin networks to moniodal 
categories, see Chapter 5.

\bibitem{lp1}C Rovelli L Smolin: Phys Rev Lett 61 (1988) 1155; Nucl 
Phys B133, 80 (1990).

\bibitem{ls-review}L  Smolin: in {\it Quantum Gravity and 
Cosmology}, eds  J  P\'erez-Mercader {\it et al}, World Scientific, 
Singapore 1992.  

\bibitem{amitaba}A. Sen, {\it On the existence of
neutrino zero modes in vacuum spacetime} 
J. Math. Phys. 22 (1981) 1781;
{\it Gravity as a spin system} Phys. Lett. B11 (1982) 89.

\bibitem{abhay1}A.A. Ashtekar, Phys. Rev. Lett. 57 (1986)
2244; Phys. Rev. D36 (1987) 1587

\bibitem{atl}A. Ashtekar, T. Jacobson and L. Smolin,
Commun. Math. Phys. 115 (1988) 631.

\bibitem{masonneuman}L. Mason and E. T. Newman,  
Commun. Math. Phys. 121 (1989) 659.

\bibitem{sdaction}T. Jacobson and L. Smolin, Phys. Lett. B196 (1987) 39;
Class. Quan. Grav. 5 (1988) 583;
J. Samuel, Pramana-J Phys. 28 (1987) L429.

\bibitem{cdj}R. Capovillia, J. Dell and T. Jacobson,
Phys. Rev. Lett. 63 (1989) 2325.

\bibitem{polyakov-loops}A. Polyakov, Phys. Lett. 82B (1979) 247;
Nucl. Phys. B 164 (1979) 171.

\bibitem{migdalmakenko}Yu. M. Makenko and A.A. Migdal,
Phys. Lett. 88B (1979) 135.

\bibitem{melattice}L. Smolin, {\it Quantum
gravity on a lattice}  Nucl. Phys. B148 (1979) 333.

\bibitem{wilson}K. G. Wilson, Phys. Rev. D 70 (1973) 2911.

\bibitem{parisi}G. Parisi,  Nucl. Phys. B100 (1975) 368; {\it On
nonrenormalizable interactions} preprint IHES/P/76/148 (1976).

\bibitem{weinberg}S. Weinberg, 
{\it General Relativity: An Einstein Centenary Survey}
 in S. W. Hawking and W. Israel,  Cambridge University Press,
1979.
eds.

\bibitem{fixedpoint}L. Smolin {\it A fixed point for quantum 
gravity}  Nucl. Phys. B208 (1982) 439.

\bibitem{ll-fractal}L.  Crane and L. Smolin, 
{\it Renormalizability of general relativity on a background
of spacetime foam}, Nucl. Phys. B267 (1986) 714;
{\it Spacetime foam as a universal regulator}, Gen. Rel. and
Grav.  17 (1985) 1209.

 \bibitem{paullee}P. Renteln and L. Smolin, Class. Quant.
Grav. 6 (1989) 275.

\bibitem{paul-diffeos}P. Renteln, Class. Quantum Grav. 7 (1990) 493-502. 

\bibitem{ezawa}K. Ezawa, Mod. Phys. Lett. A11 (1996) 349;
gr-qc/9510019; {\it Nonperturbative solutions for canonical 
quantum gravity: an overview}; gr-qc/9601050.

\bibitem{renata-lattice}R. Loll, Nucl. Phys. B444 (1995) 619;
B460 (1996)  143

\bibitem{GP-lattice}H. Fort,  J. Griego, R. Gambini, J. Pullin, 
{\it LATTICE KNOT THEORY AND QUANTUM GRAVITY IN THE LOOP
REPRESENTATION.}, gr-qc/9608033; R. Gambini and J. Pullin, 
 gr-qc/9603019 

\bibitem{tedlee}
T. Jacobson and L. Smolin, Nucl. Phys. B 299 (1988).

\bibitem{dualsuper}H.B.  Nielson, {\it  DUAL STRINGS.}
 (Bohr Inst.). NBI-HE-74-15,  
Published in Scottish Summer School 1976:465.

\bibitem{lsinbook1}L. Smolin, in {\it New Perspectives
in canonical gravity } by A. Ashtekar,
with invited contributions, Bibliopolis, Naples, 1988.


\bibitem{AI}A. Ashtekar and C. J. Isham, 
 Class and Quant  Grav 9 (1992) 1069 

\bibitem{gangof5}A Ashtekar J Lewandowski D Marlof J 
Mour\~{a}u T Thiemann:  ``Quantization of diffeomorphism
invariant theories of connections with local degrees of
freedom", gr-qc/9504018, JMP 36 (1995) 519.


\bibitem{AAinbook}A. Ashtekar, in this volume.

\bibitem{roberto1}R. De Pietri,
{\it On the relation between the connection and the loop 
representation of quantum gravity}, gr-qc/9605064.

\bibitem{lp2}R Gambini A Trias: Phys Rev D23 ,  553
(1981); Lett al Nuovo 
Cimento 38,  497 (1983); Phys Rev Lett 53,  2359 (1984); Nucl Phys 
B278,  436 (1986); R Gambini L Leal A Trias: Phys Rev D39 ,  3127
(1989);  R Gambini: Phys Lett B 255, 180 (1991)   

\bibitem{renata-volume}R. Loll, Phys. Rev. Lett.

\bibitem{viqar-int}V. Husain, B313 (1989) 711.

\bibitem{qsdi}T. Thiemann, Quantum spin dynamics I,
Harvard preprint (1996),  gr-qc/9606089.

\bibitem{qsdii}T. Thiemann, Quantum spin dynamics II,
Harvard preprint (1996),  gr-qc/9606090.

\bibitem{dpr}R. DePietri and C. Rovelli, {\it Geometry eigenvalues and
scalar product from recoupling theory in loop quantum gravity},
gr-qc/9602023, Phys.Rev. D54 (1996) 2664;
Simonetta Frittelli, Luis Lehner, Carlo Rovelli,
{\it The complete spectrum of the area from recoupling theory 
in loop quantum gravity     }
gr-qc/9608043 

\bibitem{RB}R. Borissov, Ph.D. thesis, Temple, (1996).

\bibitem{tt-length}T. Thiemann, {\it A length operator in canonical
quantum gravity} Harvard preprint 1996, gr-qc/9606092

\bibitem{tt-volume}T. Thiemann, {\it 
Closed formula for the matrix elements of the volume 
operator in canonical quantum gravity}   Harvard preprint, 1996,
gr-qc/9606091gr-qc/9601038 .

\bibitem{AL1}A. Ashtekar and J. Lewandowski, "Quantum
Geometry I: area operator" gr-qc/9602046.

\bibitem{L1}J. Lewandowski, "Volume and quantization"
gr-qc/9602035.

\bibitem{CRR}R. Borissov, R. de Pietro and C. Rovelli, preprint
in preparation, (1997).

\bibitem{carlo-grott} N Grott C Rovelli: {\it Moduli space
structure  of 
knots with intersections},   J. Math. Phys. 37 (1996) 3014,
gr-qc/9604010 

\bibitem{fm-pc}F. Markopoulou, personal communication.

\bibitem{ham1}C Rovelli L Smolin: Phys Rev Lett 72 (1994) 
446 

\bibitem{roumen-ham}R. Borissov, {\it Graphical evolution
of spin network states}, gr-qc/9605


\bibitem{instability}L. Smolin, {\it  The classical
limit and the form of the hamiltonian constraint in non-pertubative
quantum gravity} CGPG preprint, gr-qc/9609034.

\bibitem{segal}G. Segal,  {\it Conformal field theory}
oxford preprint (1988).

\bibitem{atiyah}M. Atiyah, {\it Topological quantum field theory}
Publ. Math. IHES 68 (1989) 175; {\it The Geometry and
Physics of Knots}, Lezion Lincee
(Cambridge University Press, Cambridge,1990).

\bibitem{kashdan}D. Kashdan and Lushtig, Harvard preprint.

\bibitem{kodama}H. Kodama, Phys. Rev. D 42 (1990) 2548.

\bibitem{BGP}B. Bruegmann, R. Gambini and J. Pullin,
Phys. Rev. Lett.  68,  431 (1992); Gen. Rel. and Grav. 251 (1993).

\bibitem{GP}R. Gambini and J. Pullin, 
{\it  Loops, knots, gauge theories and quantum gravity}
Cambridge University Press, 1996.


\bibitem{chopinlee}L. Smolin and C. Soo,  
{\it The Chern-Simons invariant as the natural time variable for
classical and quantum gravity}  
Nucl. Phys. B 327 (1995) 205.

\bibitem{boundary}L. Smolin, {\it Linking topological quantum
field theory and nonperturbative quantum gravity}
gr-qc/9505028,   J. Math. Phys. 36 (1995) 6417.

\bibitem{kirill1}K. V. Krasnov, {\it On statistical
mechanics of Schwarzchild black holes}.

\bibitem{meindieter}L. Smolin, {\it Time, measurement and
information loss in quantum cosmology}  in {\it Directions in
general relativity, volume 2: Papers in honor of Dieter Brill},
edited by B. L. Hu and T. A. Jacobson, Cambridge University
Press, 1993.

\bibitem{brownkuchar}D. Brown and K. V. Kuchar,
Phys.Rev. D51 (1995) 5600, gr-qc/9409001.

\bibitem{fm1}F. Markopoulou,  Class.Quant.Grav. 13 (1996) 2577,
gr-qc/9601038.

\bibitem{4dtqft} L. Crane and D. Yetter, {\it On algebraic
structures implicit in topological quantum field theories},
Kansas preprint, (1994); in {\it Quantum Topology}
(World Scientific, 1993) p. 120; 
L. Crane and I. B. Frenkel, J. Math. Phys.
35 (1994) 5136-54;

\bibitem{RR}M. Reisenberger and C. Rovelli,
{\it ``Sum over Surfaces'' form of Loop Quantum Gravity},
gr-qc/9612035.

\bibitem{bekensteinbound}J. D. Bekenstein,
Lett. Nuovo Cim 11 (1974) 467.

\bibitem{thooft}G. 'tHooft, {\it Dimensional reduction
in quantum gravity} gr-qc/9310006.

\bibitem{lenny-holo}L. Susskind, 
{\it The world as a hologram}, hep-th/9409089, J. Math.
Phys. (1995)  ; {\it Strings, black holes and Lorentz contractions}
hep-th/9308139; 
Phys. Rev. Lett.
71 (1993) 2367; Phys. Rev. D 49 (1994) 6606;
D 50 (1994) 2700;  L. Susskind and P. Griffin, {\it Partons
and black holes} hep-th/9410306;

\bibitem{fotinilee}F. Markopoulou and L. Smolin
{\it Causal evolution of spin networks}  gr-qc/9702025.
CGPG preprint (1997).

\bibitem{causalset}L. Bombelli, J. Lee, D. Meyer and
R. D. Sorkin,  {\it Spacetime as a causal set} Phys. Rev. Lett. 
59 (1987) 521.

\bibitem{louis-qc}L. Crane, {\it Clocks and categories:
Is quantum gravity algebraic?} 
J. Math. Phys. 36 (1995) 6180
in J. Math. Phys. (1996); {\it Is quantum gravity
algebraic?} in Knot Theory and Quantum  Gravity,
ed. J. Baez (Oxford University Press, Oxford).

\bibitem{carlo-relational}C. Rovelli {\it Relational
quantum mechanics},  quant-ph/9609002.

\bibitem{pluralistic}L. Smolin, {\it The Bekenstein bound,
topological quantum field theory and pluralistic quantum
cosmology}, gr-qc/9508064.

\bibitem{book}L. Smolin, {\it The Life of the 
Cosmos}, Oxford University Press (New York)
and Weidenfeld  (London) 1977, Part 5.

\bibitem{origins}L. Smolin, {\it Problems of
origins in theoretical physics}, CGPG preprint (1997).

\end{thebibliography}
\end{document}